\RequirePackage[hyphens]{url}
\documentclass[onecolumn]{article}

\usepackage[letterpaper]{geometry}
\usepackage[colorlinks]{hyperref} 
\usepackage{libertine}
\usepackage{libertinust1math}
\usepackage[T1]{fontenc}
\usepackage{natbib}
\usepackage{framed,multirow}
\usepackage{graphicx}
\usepackage{authblk}
\usepackage{amssymb}
\usepackage{latexsym}

\usepackage{url}
\usepackage{xcolor}

\usepackage{hyperref}
\usepackage{amsmath}
\usepackage{soul,color}
\usepackage{colortbl}
\usepackage{ulem,xpatch}
\usepackage{booktabs}
\usepackage{amsmath}

\DeclareMathOperator{\DSC}{DSC}
\DeclareMathOperator{\MDSC}{MDSC}
\DeclareMathOperator{\TV}{TV}

\newcommand{\mm}{\ensuremath{\text{mm}}}
\newcommand{\mednet}{\mbox{MED-Net}}

\definecolor{newcolor}{rgb}{.8,.349,.1}

\title{Segmentation of Cellular Patterns in Confocal Images of Melanocytic Lesions in vivo via a Multiscale Encoder-Decoder Network (\mednet{})}

\author[1]{Kivanc Kose\thanks{Authors contributed equally to the paper.}}
\author[2]{Alican Bozkurt\protect\footnotemark[1]}
\author[3]{Christi Alessi-Fox}
\author[4,5]{Melissa Gill}
\author[6]{Caterina Longo}
\author[6]{Giovanni Pellacani}
\author[2]{Jennifer Dy}
\author[2]{Dana H. Brooks}
\author[1]{Milind Rajadhyaksha}

\affil[1]{Dermatology Service, Memorial Sloan Kettering Cancer Center, New York, 11377,NY,USA}
\affil[2]{Electrical and Computer Engineering, Northeastern University, Boston, 02115, MA, USA}
\affil[3]{Caliber Imaging and Diagnostics, Rochester, 14623, NY, USA}
\affil[4]{Department of Pathology at SUNY Downstate Medical Center, New York, 11203, NY, USA}
\affil[5]{Skin Medical Research Diagnostics, P.L.L.C., Dobbs Ferry, 10522, NY, USA.}
\affil[6]{University of Modena and Reggio Emilia, Reggio Emilia, Italy}
\date{}
\begin{document}

\maketitle
\begin{abstract}
In-vivo optical microscopy is advancing into routine clinical practice for non-invasively guiding diagnosis and treatment of cancer and other diseases, and thus beginning to reduce the need for traditional biopsy.  However, reading and analysis of the optical microscopic images are generally still qualitative, relying mainly on visual examination. Here we present an automated semantic segmentation method called ``Multiscale Encoder-Decoder Network (\mednet{})'' that provides pixel-wise labeling into classes of patterns in a quantitative manner. The novelty in our approach is the modeling of textural patterns at multiple scales (magnifications, resolutions). This mimics the traditional procedure for examining pathology images, which routinely starts with low magnification (low resolution, large field of view) followed by closer inspection of suspicious areas with higher magnification (higher resolution, smaller fields of view). We trained and tested our model on non-overlapping partitions of 117 reflectance confocal microscopy (RCM) mosaics of melanocytic lesions, an extensive dataset for this application, collected at four clinics in the US, and two in Italy. With patient-wise cross-validation, we achieved pixel-wise mean sensitivity and specificity of $70\pm11\%$ and $95\pm2\%$, respectively, with $0.71\pm0.09$ Dice coefficient over six classes. In the scenario, we partitioned the data clinic-wise and tested the generalizability of the model over multiple clinics. In this setting, we achieved pixel-wise mean sensitivity and specificity of $74\%$ and $95\%$, respectively, with $0.75$ Dice coefficient. We compared \mednet{} against the state-of-the-art semantic segmentation models and achieved better quantitative segmentation performance. Our results also suggest that, due to its nested multiscale architecture, the \mednet{} model annotated RCM mosaics more coherently, avoiding unrealistic-fragmented annotations.
\end{abstract}
\newpage


\section{Introduction}
Many areas of medical and biological imaging have seen a recent upsurge in automated diagnosis systems using deep neural nets (DNNs). This trend is pretty much similar in many areas of traditional pathology~\citep{Litjens17,GabrieleNatMed, GooglePath}. However, the clinical application of medical imaging often involves ``edge cases'' where methods designed for natural images may not perform well. Typical challenges in these settings include large intrinsic variability, weak or inconsistent contrast, the presence of key structures in the images at distinct scales, significant class imbalance, the laborious and involved data labeling process, and the need for interpretability in terms of clinically relevant physiological features. These challenges prevent standard DNNs, even those designed for analyzing standard microscopy-based histopathological images, from achieving clinical utilization. In this work, we address one edge case of this type, analysis of morphological patterns of cellular structures in reflectance confocal microscopy (RCM) images of pigmented skin lesions. 

As we explain below, RCM has been shown to have the potential for a high impact on the assessment of such lesions and can significantly improve clinicians' ability to make accurate and reliable screening decisions on which lesions to biopsy. However, a wider adoption of RCM is hindered significantly because the images are very different visually from standard histopathology, thus making them an edge case in that context. For that same reason, automated analysis tools require solutions that go beyond standard DNN approaches and that address the challenges listed in the previous paragraph. We report here on the motivation, structure, and evaluation of a DNN architecture, which we call Multiscale Encoder-Decoder Network (\mednet{}), that was explicitly designed to overcome these edge case challenges.


Analysis of pigmented skin lesions is critical, with skin cancer being a serious medical problem worldwide. About 5.4 million new cases detected in the USA and another million in other regions (primarily parts of Europe, Canada, UK, Australia, New Zealand)~\citep{nikolaou2014emerging}. Diagnosis costs are about \$3 billion, and treatment costs another \$8 billion per year in the USA~\citep{SkinCancerCost}. RCM is an emerging non-invasive optical diagnostic tool based on examination of living tissue morphology directly on patients, on the fly, and at the bedside or in the clinic. After more than two decades of development and translation, \textit{in vivo} RCM is advancing into clinical practice for non-invasively guiding diagnosis and treatment of cancer~\citep{rajadhyaksha2017reflectance}. RCM imaging, combined with the current clinical standard for visual examination, known as dermoscopy, reduces the benign-to-malignant biopsy ratio by about a factor of two compared to dermoscopy alone~\citep{Alarcon14,Pellacani14,Pellacani16,Borsari16}.

Although RCM images have a $\mu$m-level resolution like standard histopathology, their appearance is quite different because they are collected \textit{in vivo}. One difference is that the images are acquired in an en face orientation, as opposed to the ``vertical'' (i.e.\nobreak\ normal to the skin surface) sections typically used in the pathology of excised specimens. Another is that, due to lack of \textit{in vivo} contrast agents, images have only one source of contrast, reflectance, and therefore are displayed in grayscale, whereas standard H\&E pathology is in color contrast (the purple and pink appearance). Instead of color contrast, skin and cellular structures are differentiated by intricate multiscale textural patterns in RCM images.

Diagnosis of melanocytic lesions using RCM is primarily based on the identification of four cellular morphological patterns in RCM mosaics acquired at the dermal-epidermal junction (DEJ). These mosaics typically span rectangular-shaped areas with 4\nobreakdash-6\nobreak\;mm at one side~\citep{ScopeTermChap17}. The patterns in the mosaics are composed of heterogeneous cellular formations, appear at highly varying scales with highly varying shapes, and with diffused transition boundaries in between. Moreover, the images are contaminated by intrinsic speckle noise. All these aspects are characteristic of high-resolution optical microscopy \textit{in vivo}. 

These characteristics present challenges for human readers who are trained extensively to interpret H\&E pathology. Learning to read and perform a qualitative examination of RCM images demands significant effort and time for novices, and results tend to be highly subjective, with high levels of inter-reader variability even among experts. The steep learning-curve and large inter-reader variability have become a significant impediment to broader RCM adoption by clinicians, which strongly motivates the development of automated computational tools for both clinical guidance and clinical training. 

Existing medical image segmentation applications are developed for identifying target structures that typically have
\begin{enumerate}
   \setlength\itemsep{0em}
    \item predefined shapes with noticeable boundaries (e.g.\nobreak\ organs~\citep{CT-MR-FCN,Prostate2017}, cells~\citep{ronneberger2015u,unet19}),
    \item distinct contrast compared to the background (e.g.\nobreak\ cells, retinal vessels~\citep{FuRetina16}),
    \item predefined spatial location within the view (e.g.\nobreak\ organs, retinal layers~\citep{Gu19}, lesions~\citep{ISIC2016,ISIC2017}).
\end{enumerate}
On the other hand, the morphological structures encountered in RCM images are complex in shape, have ambiguous boundaries, vary in size, change appearance under inherent speckle noise, and appear at arbitrary spatial locations within the field of view. Therein our experience has convinced us that neither the existing semantic segmentation approaches developed for other medical imaging modalities~\citep{ronneberger2015u,unet19,CT-MR-FCN,Prostate2017,ISIC2016,ISIC2017} nor the existing very dewep neural network architectures~\citep{segnet,deeplab} can be effectively used for RCM mosaics. These models contain very large numbers of parameters to optimize, making them prone to overfitting with the type of limited and class-imbalanced training data available for RCM. Moreover, in deep network architectures with limited training data, the training of the layers which are farther away from the output is challenging as the partial derivatives that define the coefficient updates tend to get smaller as the error propagates from the output towards the input layers. 

To respond to these particular challenges of automated analysis of RCM images, we developed a multiscale neural network called \mednet{} for semantic segmentation of textural patterns in segmented lesions, based on the morphological patterns that have been defined by expert RCM readers. The architecture of \mednet{} was driven by two key observations about clinical practice. First, our multiscale structure was inspired by the typical procedure for examining pathology in RCM mosaics clinically, which routinely starts with low magnification and low resolution in a large field of view (2X-4X, \mbox{$\sim$1-5 $\mu$m/px}, over \mbox{5-10 mm}) followed by closer inspection of suspicious areas with higher magnification and higher resolution in smaller fields of view (10X-40X, \mbox{0.2-1.0 $\mu$m/px}, over \mbox{0.5-2 mm}), and then often returns to lower magnification to integrate features found at higher resolution into a broader semantic setting. \mednet{} models textural patterns at multiple scales (magnifications, resolutions), starting from a coarse scale and proceeding to finer scales. Semantic segmentation at each scale is handled by subnetworks, which are fully convolutional encoder-decoder neural networks capable of generating label maps at the same scale as their input. The capacity (number of layers and coefficients) of the subnetworks depends on the complexity of the segmentation task at the given scale (e.g.\nobreak\ coarser scales use smaller subnetworks as there is less detail at those scales). Consecutive subnetworks in the multiscale hierarchy explicitly cooperate, leveraging the correlation across scales. Each subnetwork utilizes the encoded feature representation (called the bottleneck representation) from the immediate predecessor subnetwork by integrating it into its feature representation at the equivalent level. 

Similarly, the semantic segmentation estimation of each subnetwork is used as a prior in the subnetwork at the finer scale, so that each subnetwork only refines the coarser-scale estimates rather than solving the whole segmentation problem from scratch. However, using several subnetworks in a cascaded fashion makes the model rather deep and can make training difficult. To solve this problem, we employ a method called ``deep supervision''~\citep{zhu2017deeply}. We compare the output of the subnetwork at every scale against ground truth segmentation downsampled to the same scale. This supervision gives us direct access to deeper layers (early subnetworks) and allows efficient updates to avoid vanishing gradients during training. 

Second, we use a set of four cell-morphological patterns (textural structures) that have been identified by clinicians~\citep{ScopeTermChap17} along with two ``extra" classes for artifacts and non-lesion background. Rather than designing a binary classifier to simply classify lesions as suspicious or non-suspicious, we aim to respond clinicians' need for transparency in diagnostics by providing them a scheme that reports more finely grained results in this ``edge cases" setting. Indeed, given this critical need for transparency and its intrinsic advantage for both rapid reader throughput and education, it is of critical importance to generate pattern class masks rather than just binary classifications. Similarly, we chose pixel-wise instead of image-wise classification, because in the latter, the clinician only has access to the final diagnostic prediction, while pixel-wise segmentation reports the spatial location of the diagnostic findings, making the diagnostic process more interpretable. 







The precursor to \mednet{}, named MUNet, was developed as a feasibility study~\citep{KoseM18}. Here we significantly extend MUNet in the following ways:
\begin{enumerate}
    \setlength\itemsep{0em}
    \item MUNet only provides feedback between consecutive layers via output label maps, whereas \mednet{} also shares feature representations between consecutive subnetworks (Fig.~\ref{fig:arch}, Section~\ref{sec:Arch}).
    \item We trained \mednet{} using a novel loss function that incorporates a total variation constraint to regularize the smoothness of the output label maps (Section~\ref{LossFunc}).
    \item We greatly expanded the dataset used to train and test \mednet{} compared to MUNet, using what is, in the RCM context, an unprecedentedly rich set of labeled data, 117 mosaics, collected at six different clinics in the US (4) and Italy (2). In addition to only having more data available, here we were able to carry out cross-validation with data stratified by clinic-of-origin, providing a more realistic prediction of future performance. We note that while in the context of DNNs, this is a rather small dataset, it is large for RCM due to the difficulty of labeling, an aspect of the ``edge case'' nature of this problem. 
\end{enumerate} 
Labeling datasets is laborious and challenging, even for experts. Indeed, only $58\%$ of the pixels in the dataset were labeled by our experts due to these difficulties. Thus, another feature of \mednet{} is the ability to train on ``partially-labeled'' data, where only arbitrarily-shaped parts of training images are labeled, but be capable of classifying full images. In our quantitative evaluations, we can only compare to the labeled pixels as we only have ground-truth there, but we show our visual segmentation results on the full images (Fig.~\ref{fig:ResultsSeg}). We evaluated the segmentation performance of \mednet{} using the Dice coefficient, as well as the sensitivity and the specificity of the model in identifying the patterns. We compared \mednet{} results against 4 well-known DNN models (FCN~\citep{fcn}, SegNet~\citep{segnet}, DeepLab~\citep{deeplab} and UNet~\citep{ronneberger2015u}). 


In the following sections, we discuss the design of \mednet{} in detail, explain the algorithmic choices we made to overcome unique issues encountered in semantic segmentation of \textit{in vivo} microscopy images, and present the results of our tests on mosaics of melanocytic skin lesions. 

\section{Materials and Methods}
Our study set is composed of 117 RCM mosaics of melanocytic skin lesions collected at the DEJ level. 31 of these mosaics were acquired at 4 different clinics in the US (Memorial Sloan Kettering Cancer Center (New York, NY), University of Rochester (Rochester, NY), Loma Linda University Health (Loma Linda, CA), and Skin Cancer Associates (Plantation, FL) ) and the other 86 at clinics at the University of Modena and Reggio Emilia (Italy). All mosaics were collected under the required IRB (USA) and Ethics Committee (EU) approvals and de-identified (patient metadata was removed). The study set was chosen to reflect the data diversity encountered in daily clinical practice. At each clinic, the imaging was carried out with a commercial confocal microscope (Vivascope 1500, Caliber I.D.) with a spatial resolution of $0.5\;\mu \text{m/px}$. Mosaic sizes varied from $7000 \times 8000 $ pixels up to $12000 \times 12000$ pixels, corresponding to an area between 14 and 36 $\mm^2$. The size of the mosaics was determined by the clinical need to be able to evaluate the cellular morphological patterns that characterize melanocytic lesions accurately. 


\begin{figure*}[ht]
    \centering
    \includegraphics[width=0.8\linewidth]{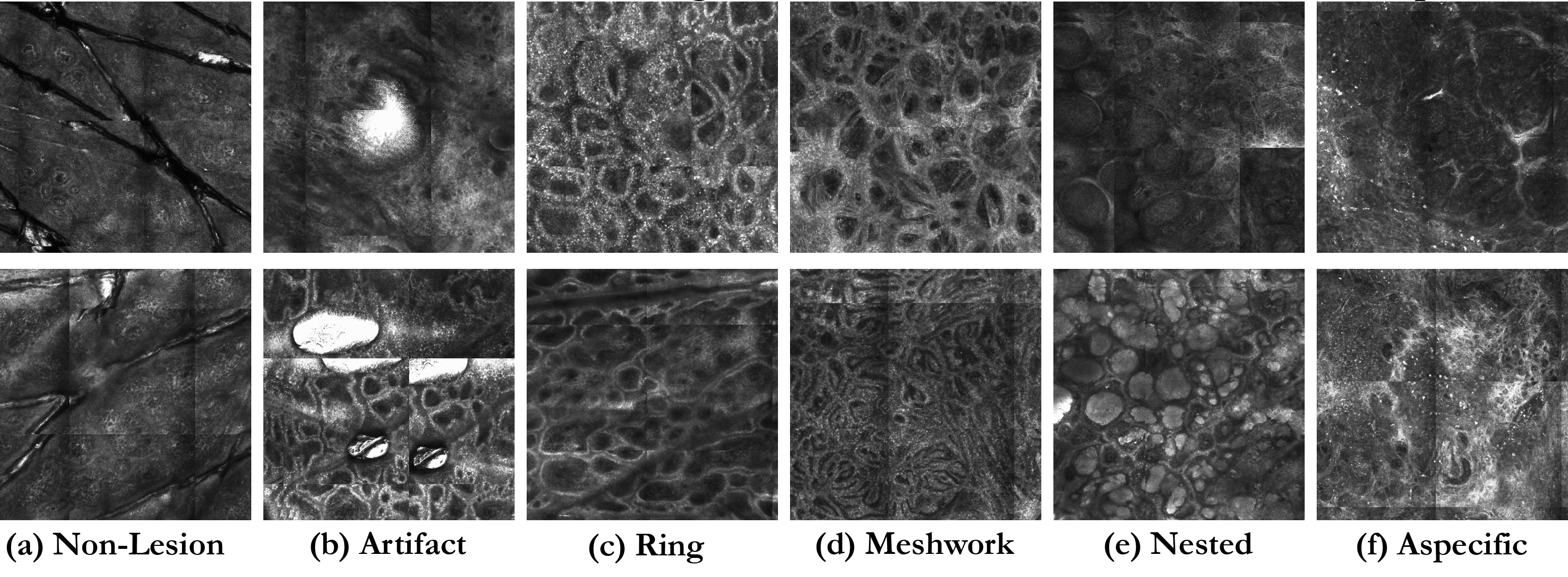}
    \caption{Two examples for each of the six distinct patterns (four cellular morphological and two other patterns). as seen in reflectance confocal mosaics at the dermal-epidermal junction in melanocytic skin lesions.}
    \label{fig:patterns}
\end{figure*}

We set as our goal the segmentation of these mosaics into six clinically important classes. Four of them are cellular morphological patterns,i.e.\nobreak\ ring, meshwork, nested, and aspecific. These patterns are routinely observed in RCM mosaics of melanocytic neoplasm collected at the DEJ~\citep{ScopeTermChap17}. We added two additional classes for non-lesion areas and areas dominated by imaging artifacts~\citep{Artifacts}, leading to six total classes in our segmentation task.\footnote{Regions that were beyond the lesion borders or within the lesion but not at the DEJ were classified as non-lesion. Artifact regions and their imaging characteristics include saturation (too bright), under-illumination (too dark). Air bubbles in the index-matching oil (appear as bright highly reflecting blobs or amorphous structures) or in the immersion media (appear as large dark round/oval areas) can also obscure the underlying tissue morphology, resulting in artifacts in the images~\citep{Artifacts}.} Exemplars of these six classes are shown in Fig.~\ref{fig:patterns}. 

\begin{figure}[ht]
    \centering
    \includegraphics[width=0.65\linewidth]{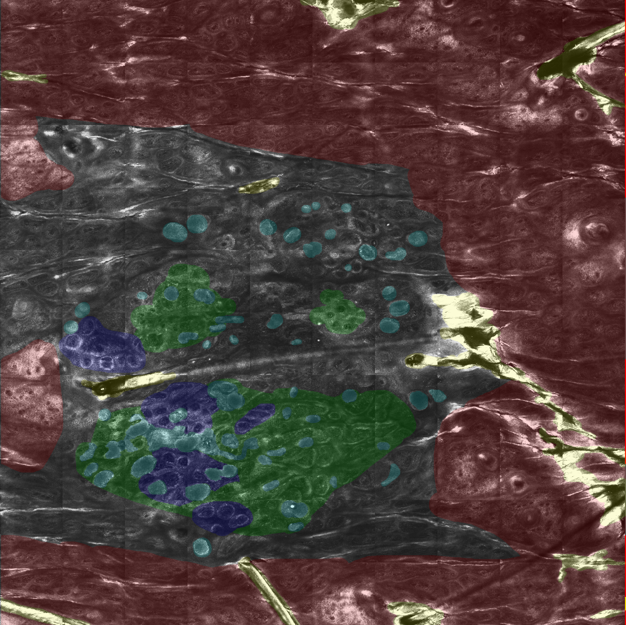}
    \caption{An example mosaic and its corresponding expert labeling. Colors indicate the labels; Red: Non-Lesion, Yellow: Artifact, Green: Meshwork, Blue: Ring, Cyan: Nested. Grey colored areas are not labeled, and are ignored in training and quantitative evaluation.}
    \label{fig:exampleSeg}
\end{figure}

Ground truth maps for these six classes came from labels determined by the consensus of 2 expert readers (co-authors MG and CAF), labeled using the open-source software package Seg3D (University of Utah,~\citep{Seg3D}). Labeling was conducted in a non-exhaustive manner, meaning that pixels not labeled as any of the six classes were given a distinct ``ignore" label. Pixels were not labeled either because the distinction between the labels was not clear due to the existence of mixed patterns or because they would have required excessive time and effort to label, in the readers' judgement. Overall, $58\%$ of the pixels were labeled (Table~\ref{tab:DataStats}). We show a sample labeled mosaic in Fig.~\ref{fig:exampleSeg}. The unlabeled portions of the mosaics were omitted during both training and quantitative testing. However, the readers qualitatively assessed the algorithm's segmentations even for these unlabeled regions. The distribution balance of the six labels over the whole dataset is given in Table~\ref{tab:DataStats}.

\subsection{Semantic Segmentation Network Architecture}\label{sec:Arch}
\mednet{} is composed of multiple encoder-decoder subnetworks nested together (Fig.~\ref{fig:arch}). Each subnetwork processes the input image starting at a specific scale and outputs a segmentation map at the same scale. To the best of our knowledge, \mednet{} is different from existing networks in the following aspects. In similar existing approaches~\citep{refinenet,Jiang18,Islam17,deeplab,pspnet, Fu18, Zhou18, Feng18,Sarma18, Fu19}, the subnetworks are cascaded so that they share only features across networks, or else they independently solve the same segmentation problem and then, only at the end, fuse the results. More similar to \mednet{}, \citet{Fergus15} use three separate networks to process the input images at different scales in a cascaded manner resembling our approach. They feed the output of subnetworks into the input of the following subnetworks, so the individual models provide feedback to each other. However, in their approach, due to lack of feedback at the individual subnetwork level (e.g.\nobreak\ deep supervision~\citep{zhu2017deeply}), the output of each subnetwork is not final output (e.g.\nobreak\ in their case, a depth map) at respective scale, but a feature representation. Unlike all these approaches, \mednet{} shares intermediate results in two ways. It shares the segmentation outputs across subnetworks (Fig.~\ref{fig:arch}) by using them as a prior that becomes part of the input for subsequent subnetworks. Through the use of deep supervision~\citep{zhu2017deeply}, the output of each subnetwork is compared against a ground truth segmentation and forced to be an intermediate label prediction at the given scale it operates. 

Moreover, \mednet{} also shares feature representations between matching levels of consecutive subnetworks. These subnetwork interconnections are not present in previous approaches~\citep{refinenet,Jiang18,Islam17,deeplab}. Backpropagating the final loss through the network can lead to inefficient training of the layers that are farther from the output. Therefore, to effectively train the individual subnetworks, we provide direct feedback to them, a method known as deep supervision~\citep{zhu2017deeply}. Overall, sharing intermediate feature representation, using intermediate label predictions as priors, and deep supervision to individual subnetworks are the three main innovations in the \mednet{} architecture.

The elementary units of subnetworks in \mednet{} consist of residual blocks~\citep{HeNormal}, which are generally concatenations of convolutions, non-linearities, and batch normalizations. Downsampling is carried out through non-unity stride of the first residual block, and upsampling is applied to processing block outputs. The sequence of downsampling processing blocks (encoder) is followed by a sequence of upsampling processing blocks (decoder). Thus if we had a single scale, the architecture would be very similar to a Fully Convolutional Network (FCN32)~\citep{fcn} with encoder-decoder topology. However, here we have $M$ subnetworks that solve the segmentation problem starting from a different scale of the input image. Subnetworks in this cross-scale hierarchy share information (feature representations) directly through skip connections from bottleneck representations of their predecessor scale subnetwork. This information exchange is done via multiplication of tensor representations at comparable scales to act like attention mechanisms~\citep{SqueezeExcite}. Also, the output segmentation probability map (a vector of six probabilities per pixel) at each scale (except the finest) is upsampled and then concatenated with the original or directly downsampled image at the next finer scale and used as the input for the subnetwork at that next scale. More precisely, let $I^0$ and $L^0$ be the original image and corresponding ground truth labeled image, and $I^m$ and $L^m$ be those images after $2^m$ times downsampling in both spatial dimensions ($m=0,\dots,M-1$). The subnetwork at the coarsest scale takes only $I^{M-1}$ as input and produces a probability map $\hat{L}^{M-1}$, which represents the likelihood of each pixel belonging to a particular class. For all other subnetworks (i.e.\nobreak\ $m\in\left[0,M-2\right]$), we fuse the segmentation coming from subnetwork $m+1$ ($\hat{L}^{m+1}$) with the level $m$ version of the input ($I^{m}$) via concatenation. The final segmentation probability map is $\hat{L}^{0}$, which is at the same resolution as the input image of the overall model. 

The subnetwork depth parameter $M$ is a design choice, and one can also vary the scale factor between subnetworks, which we set to 2, leading to a 3-level version of \mednet{}. Likewise, the scale difference of the input between consecutive levels is another design choice and can be determined according to needs and computational capabilities. In addition, the overall architecture is modular in the sense that one can replace our subnetwork architecture (including a different design of the processing blocks) with any other relevant subnetwork architecture and then assemble a \mednet{} version of that network.


Each \mednet{} subnetwork for $m>0$ has the same architecture as the subnetwork at scale $m-1$ but with two additional blocks: One encoder block before the bottleneck feature representation and one deconvolution block at the input of the decoder. Note that the weights in each corresponding block differ across subnetworks; weights are not shared between layers. Information is shared between subnetworks only through the skip connections described above. 

\subsection{Loss function}\label{LossFunc}
\begin{figure*}[ht]
    \centering
    \includegraphics[width=\textwidth]{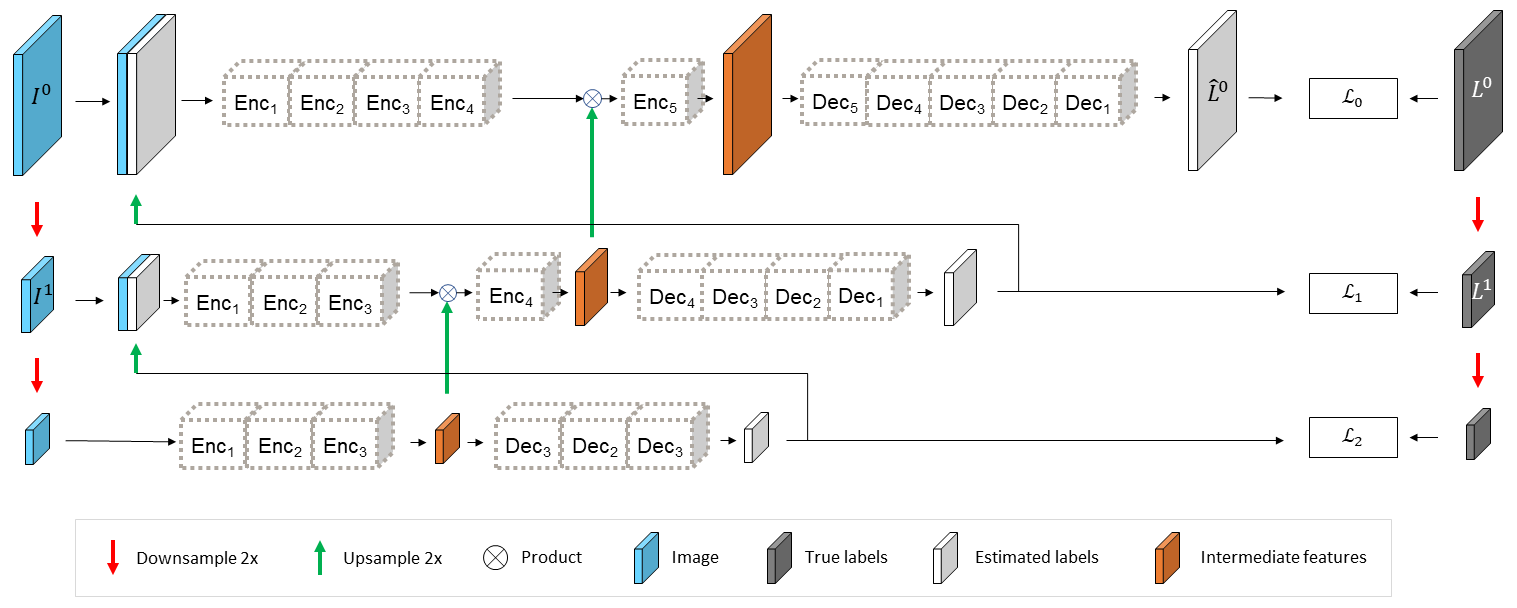}
    \caption{Our architecture is composed of 3 nested fully convolutional networks that generate semantic segmentation at different scales. Red arrows denote 2x downsampling, and green arrows denote 2x upsampling. Output segmentations at lower magnifications are fed into the next level via concatenation. The loss at each level (scale) is calculated and backpropagated for deep supervision of the subnetworks.}
    \label{fig:arch}
\end{figure*}

The loss function was designed to take three distinct factors into account:
\begin{enumerate}
    \setlength\itemsep{0em}
    \item Appropriateness of segmentation (e.g.\nobreak\ generating labels that change smoothly across the image).
    \item Ability to handle imbalances in label distribution of the training data.
    \item Applicability to multiclass labeling. Thus we used a modified version of the soft-Dice loss calculated between $\hat{L}^{m}$ and $L^{m}$ (see Fig.~\ref{fig:arch}) at each level.
    \end{enumerate}
The standard Dice Similarity Coefficient $\DSC(A,B)={2|A \cap B|}/{\left(|A|+|B|\right)}$~\citep{dice1945measures} is commonly used for binary segmentation and is known to be robust against label imbalance in the data. In its original binary formulation, DSC explicitly represents only true-positive samples, while true-negative cases are automatically optimized simultaneously. However, similar to~\citet{Salehi17}, we found that directly extending this formulation to the multilabel case by treating each label as a binary classification task did not put enough emphasis on true-negatives samples. Therefore, we modified the soft-Dice coefficient also to consider true-negative samples in the loss calculation, as described next.

Suppose we have $W\times H \times K$ sized tensors $L^m$ and $\hat{L}^m$, where $L^m$ is one-hot encoded ground truth at the subnetwork level $m$. The entries $L^m_{ij}=\mathbf{e}_k$ if pixel $\left(i,j\right)$ is labeled as class $k$, where $\mathbf{e}_k$ is a one-hot vector of length $K$ with 1 in its $k^{th}$ entry and 0 everywhere else. $\hat{L}^m$ is the neural network output, such that at each $(i,j)$ pixel, $\hat{L}^m_{ijk}\in\left[0,1\right]$ and $\sum_k \hat{L}^{m}_{ijk}=1$. Our modified loss function is:
$$
\MDSC(L^m,\hat{L}^m) = \sum_{k=0}^{K-1} \left(1-\frac{2{\sum_{i,j}L^m_{ijk}\hat{L}^m_{ijk}}}{{{\sum_{i,j}({L^m_{ijk}})^2+({\hat{L}^m_{ijk}})^2}+\epsilon}}\right)
$$
$$
\quad +\sum_{k=0}^{K-1} \left(1-\frac{2{\sum_{i,j}\left(1-L^m_{ijk}\right){\left(1-\hat{L}^m_{ijk}\right)}}}{{{\sum_{i,j}\left(1-L^m_{ijk}\right)^2+{\left(1-\hat{L}^m_{ijk}\right)}^2}+\epsilon}}\right)
$$
where $\epsilon$ is a small value in order to avoid division by zero. The first part of the equation is the standard soft-Dice loss, which encourages agreement between true positive labels, while the second part of the equation also encourages agreement between true negative predictions. 
To ensure smoothness of the prediction label map and avoid small isolated segmentation labels, we regularize the loss function using the total variation (TV) of the output label map.
$$
    \TV(\hat{L}^m)=\sum _{{i,j,k}}\left|\hat{L}^m_{{i+1,j,k}}-\hat{L}^m_{{i,j,k}}\right|+\left|\hat{L}^m_{{i,j+1,k}}-\hat{L}^m_{{i,j,k}}\right|.
$$
Combining MDSC and TV losses, the loss applied at each subnetwork level is
$$\mathcal{L}_m= \MDSC(L^m,\hat{L}^m)+\gamma \TV(\hat{L}^m).
$$ 
We set the regularization parameter empirically, $\gamma=10^{-6}$, which kept the total variation cost to $[0.1,0.01]$ of the soft-Dice loss. In our experiments, we observed that keeping the total variation cost within this range of the soft-Dice loss provided a good balance between smoothness and the accuracy of produced label maps. 

As shown in Fig.~\ref{fig:arch}, we calculate $\mathcal{L}^m$ between outputs of each subnetwork and the label map at the respective scale for each scale $m$, and the overall loss as the sum of losses across all subnetworks/scales $\mathcal{L}= \sum_{m=0}^{M-1}\mathcal{L}^m$. Doing so, we effectively gain direct access to the deeper layers of the network, as is done with deep supervision~\citep{zhu2017deeply}. However, the subnetworks are not trained disjointly as they are connected via skip connections, resulting in joint optimization of all subnetwork parameters.

\subsection{Implementation Details}
In this section, we discuss specific parameter choices in our implementation of \mednet{} on RCM mosaics. These choices were made to fit available hardware resources (e.g.\nobreak\ GPU memory, number of GPUs) and problem characteristics (e.g.\nobreak\ data sampling and augmentation scheme). We report them so that readers can replicate our work, and we also anticipate that they will provide a guideline towards applying this structure to other segmentation problems. 

Before training the \mednet{} model, we needed to make two important choices regarding; (i) the resolution of the mosaics to be processed and (ii) the size of the input images to the network. Although the network architecture can segment arbitrarily sized images, we processed the RCM mosaics in patches (portions of the mosaic) due to memory limitations of the GPU we used. Note that the patches needed to be larger than $2^4$ pixels per dimension because we used 2-strides (effectively downsampling by 2) at least at 4 levels of encoder blocks. To determine useful patch-sizes, we consulted our expert readers, who reported that in their experience, the morphological patterns of interest could still be reliably identified at 2 $\mu$m/px resolution, 4-times lower than that of the RCM acquisition system. Thus before feeding the mosaics to \mednet{}, we downsampled them by 4. The readers also reported that a $0.5\;\mm\times0.5\;\mm$ field of view is typically large enough to identify these same patterns reliably. Thus we processed the mosaics in patches of $256\times 256$ pixels after downsampling. 

All models are trained using the same training parameters. We trained each model for 200 epochs, using a base learning rate of 0.01, batch size of 48, and weight decay of $10^-8$. We exponentially decayed the learning rate to one-tenth of the base value throughout the training. For a fair comparison, we kept the number of trainable parameters for all networks at 6 million. All the convolutional layers are initialized with He Normal initialization~\citep{HeNormal}.

We also implemented data augmentation through spatial sampling. In order to cover all possible patches that could be extracted from the mosaic, we devised the following patch extraction procedure. Before each epoch, we extract $512\times 512$ pixels patches in a sliding window fashion with a $50\%$ overlap. Then, at each epoch of training, we extracted $256\times 256$ pixel patches at random locations within the larger patches. 

In order to account for inevitable variations during RCM image acquisition, such as changes in laser power (illumination intensity), distortion in tissue, speckle noise, and the orientation of the microscope, we applied data augmentation on the extracted patches. At each epoch, we 
\begin{enumerate}
    \setlength\itemsep{0em}
    \item rotated each patch at a random angle up to 180 degrees
    \item randomly flipped the patch horizontally and vertically,
    \item added a random intensity value in [-20, 20]\footnote{At the augmentation stage, the pixels values are in the range [0,255] so we clipped the added intensity for the brightest pixels.}
    \item zoomed in/out randomly up to 10\%, \item randomly sheared the patches ($\theta = 0.2$),
    \item added signal-dependent Gaussian-distributed pseudo-speckle noise (with uniform random multiplication parameter of 0.2). 
\end{enumerate}

During inference, the output of the networks is six probability maps, one for each label (represented as a $256\times 256\times 6$ tensor) over a 0.5 $\mm^2$ field of view. Due to the use of padded convolutions, the network produces less reliable segmentation results at the borders of the patches. To compensate, we extracted and processed patches in an overlapping fashion, resulting in multiple soft decisions for each pixel. Specifically, we extracted patches at a stride of 32 pixels, leading to up to 8 different decisions per pixel. We then weighted each patch's probability map for each label with a spatial Gaussian mask whose variance was half of the patch size before summing the overlapping probability maps. Finally, we chose the class with the highest resulting probability for each pixel. 

\section{Results}
\label{sec:Results}
We report the results of testing on two distinct training scenarios. In Scenario~1, we pooled data across all sites, then stratified by the patient for training, validation, and testing (5-fold stratified cross-validation). In Scenario~2, we first stratified by clinics, only used the data from clinics in Europe for training and validation, and then tested only on data from the US. The validation set was used to probe the performance of the model throughout training, and the test set was used to evaluate the performance of the trained models quantitatively. We chose to train on the European data and test on the US data, and not vice-versa, both due to the limited size of the US data set and also because the US data came from a larger number of clinics, thus better mimicking a more realistic application scenario. Results from the first scenario are described in Section~\ref{PatientExp} and results from the second scenario in Section~\ref{ClinicExp}. Each fold used in Scenario~1 is also stratified by the class label in the training/test split to ensure a representative sampling of training data in the face of the class imbalance in our data. Specifics of the data distribution over the training, validation, and test sets for both scenarios are given in Table~\ref{tab:DataStats}.

In addition to \mednet{}, we also tested 4 other widely used deep segmentation networks; FCN~\citep{fcn}, SegNet~\citep{segnet}, DeepLab~\citep{deeplab}, and UNet~\citep{ronneberger2015u} for comparison purposes. To try to ensure fair comparisons, we used a similar number of trainable parameters in each network ($\sim 6\times 10^6$). All of the networks were trained using similar training parameters (e.g.\nobreak\ learning rate, weight decay, batch size) for 200 epochs using the MDSC+TV loss described above.


\begin{table}[htbp]
  \centering
  \caption{Class distribution statistics: The top portion reports the distribution of labels for both scenarios. In Scenario~1, we were able to balance distribution across training and test sets to within 1\% (stratified cross-validation). Class distributions in training and test sets are explicitly given for Scenario~2. In the bottom portion, we report on the size of the datasets in terms of both images and labeled pixels, as well as on the overall fraction of pixels that were labeled.}

\begin{tabular}{rccc}
\toprule
& Scenario~1 & \multicolumn{2}{c} {Scenario~2} \\
\cmidrule(l{2pt}r{2pt}){2-2} \cmidrule(l{2pt}r{2pt}){3-4}
& Whole Dataset & Europe (Train) & US (Test) \\
\midrule
Background & 17\% (83.7M) & 19\% (65.3M)& 13\% (18.4M) \\
Artifact & 19\% (92.6M)   & 20\% (67.7M)& 18\% (24.5M)\\
Mesh & 20\% (97.5M)       & 21\% (73.7M) & 17\% (23.8M)\\
Nest & 5\% (25.6M)        & 6\% (19.0M) & 5\% (6.6M)\\
Ring & 28\% (136.6M)      & 23\% (78.8M) & 40\% (57.8M)\\
Aspecific & 10\% (50.5M)   & 12\% (40.3M) & 7\% (10.1M)\\
\midrule
labeled Pixels & 57\% (486.6M) & 51\% (344.7M) & 60\% (141.8M) \\
\# labeled Images & 117 & 86 & 31 \\
\bottomrule
\end{tabular}
  \label{tab:DataStats}%
\end{table}%

\subsection{Scenario-1: Patient-Wise Cross-Validation Experiment}\label{PatientExp}
As described above, in this scenario, we ``patient-wise partitioned" the dataset into 5 stratified folds, meaning that each fold contained similar proportions of class labels. Training, validation, and test sets approximately corresponded to 70, 10, and 20 percent of the data in each fold, respectively.  

In Table~\ref{tab:ResultsCV}, we present the segmentation performance of all four networks for Scenario~1 in terms of sensitivity, specificity, and the Dice coefficient. On average, \mednet{} modestly outperforms the other networks in terms of sensitivity (by $0.02$ to $0.12$), although the comparison differs across classes. On specificity, all four networks perform similarly both on average and by class. The Dice coefficient values are consistently better for \mednet{} than the compared methods except for FCN on the Nest class. In general, FCN was the closest to \mednet{}. 

A closer comparison of the model output with ground truth labels revealed that in general, all models confused the meshwork class with the ring and aspecific classes. This result is interesting, because anecdotally we have been told that novice clinicians also suffer from the same problem due to the wide range of variations in the appearance of the meshwork pattern. Moreover, visual examination of the results by our experts confirmed that most of the falsely classified meshwork pattern samples contain "deformed" variations of the pattern, which they reported are typically also misclassified by novice readers. 

\begin{figure*}[htbp]
    \centering
    \includegraphics[width=\textwidth]{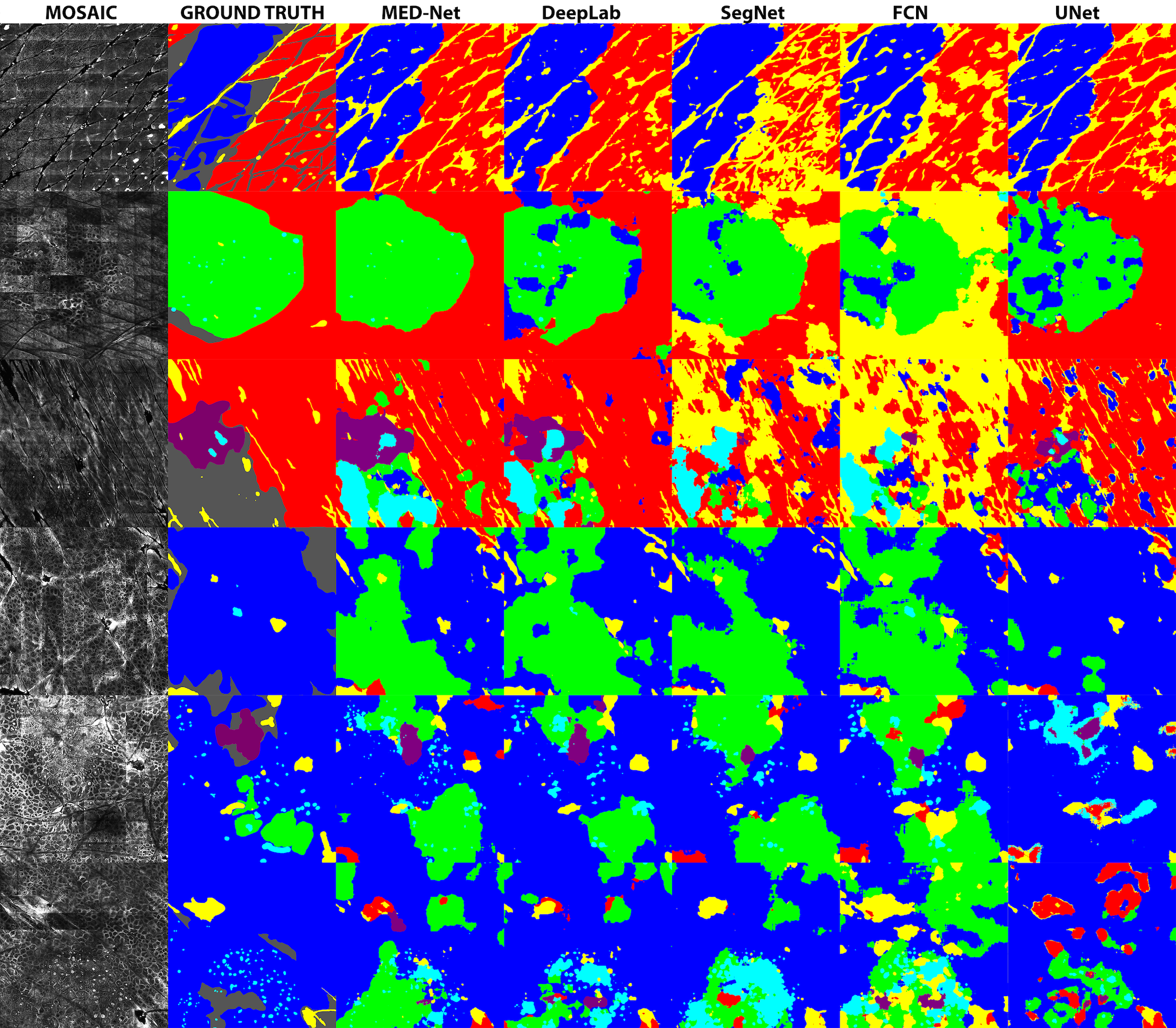}
    \caption{Example segmentation results of 6 mosaics for Scenario~1. Color scheme is the same as used in Fig.~\ref{fig:exampleSeg}. The ground truth segmentations are compared to the outputs of \mednet{} and other state-of-the-art-methods. Images are not exhaustively annotated by the readers. Pixels that are not annotated (dark grey label) are ignored during training. During the testing phase, these pixels are discarded from sensitivity and specificity calculations.}
    \label{fig:ResultsSeg}
\end{figure*}

To obtain a qualitative assessment of \mednet{} outputs, we presented the segmentation maps produced by \mednet{} to our experts. In particular, we asked them to review the automated annotation of the algorithm over the ``unlabeled areas". Their qualitative assessment of the results was very positive and confirmed that the model performed very well in annotating most of the unlabeled areas in the mosaics. We show an example in Fig.~\ref{fig:ResultsSeg}. The gray-colored areas in the figure represent the unlabeled areas. \mednet{} typically extended the labels of the neighboring labeled areas over the unlabeled sections, providing smoother label maps than the other methods. 

\begin{table*}[htbp]
  \centering
  \caption{Results for Scenario 1 Patient-Wise. The best results for each metric and label are highlighted in bold.}
    \begin{tabular}{rccccc}
\toprule   
& \multicolumn{5}{c}{Sensitivity} \\
\cmidrule(lr){2-6}          
& MED-Net & DeepLab & SegNet & FCN & UNet \\
\midrule
Background & 0.83 $\pm$ 0.05 & 0.91 $\pm$ 0.05 & \textbf{0.94  $\pm$ 0.02} & 0.86 $\pm$ 0.04 & 0.85 $\pm$ 0.11 \\
Artifact & \textbf{0.83 $\pm$ 0.10} & 0.58 $\pm$ 0.18 & 0.57 $\pm$ 0.18 & 0.81 $\pm$ 0.08 & 0.78  $\pm$ 0.11 \\
Meshwork & \textbf{0.59 $\pm$ 0.09} & 0.47 $\pm$ 0.10 & 0.45 $\pm$ 0.14 & \textbf{0.59 $\pm$ 0.08} & 0.22 $\pm$ 0.15 \\
Nest & 0.59 $\pm$ 0.13 & 0.49 $\pm$ 0.22 & 0.53 $\pm$ 0.16 & \textbf{0.60 $\pm$ 0.14} & 0.34 $\pm$ 0.20 \\
Ring & 0.85 $\pm$ 0.07 & 0.82 $\pm$ 0.10 & 0.74 $\pm$ 0.12 & 0.82 $\pm$ 0.14 & \textbf{0.87 $\pm$ 0.08} \\
Aspecific & \textbf{0.53 $\pm$ 0.22} & 0.21 $\pm$ 0.32 & 0.38 $\pm$ 0.031 & 0.40 $\pm$ 0.28 & 0.48 $\pm$ 0.28 \\
\toprule
& \multicolumn{5}{c}{Specificity} \\
\cmidrule(lr){2-6} 
& MED-Net & DeepLab & SegNet & FCN & UNet \\
\midrule
Background & \textbf{0.99 $\pm$ 0.01} & 0.84 $\pm$ 0.09 & 0.81 $\pm$ 0.14 & 0.97 $\pm$ 0.01 & 0.96 $\pm$ 0.01 \\
Artifact & 0.92 $\pm$ 0.02 & \textbf{0.96 $\pm$ 0.02} & 0.94 $\pm$ 0.03 & 0.94 $\pm$ 0.01 & 0.89 $\pm$ 0.04 \\
Meshwork & 0.91 $\pm$ 0.02 & 0.92 $\pm$ 0.04 & 0.92 $\pm$ 0.06 & 0.89 $\pm$ 0.05 & \textbf{0.96 $\pm$ 0.03} \\
Nest &\textbf{ 0.99 $\pm$ 0.01} & \textbf{0.99 $\pm$ 0.01} & \textbf{0.99 $\pm$ 0.00} & \textbf{0.99 $\pm$ 0.01} & \textbf{0.99 $\pm$ 0.02} \\
Ring & 0.88 $\pm$ 0.06 & 0.087 $\pm$ 0.04 & \textbf{0.92 $\pm$ 0.02} & 0.86 $\pm$ 0.05 & 0.81 $\pm$ 0.04 \\
Aspecific & 0.98 $\pm$ 0.01 & \textbf{0.99 $\pm$ 0.00} & 0.97 $\pm$ 0.02 & \textbf{0.99 $\pm$ 0.01} & 0.95 $\pm$ 0.05 \\
\toprule
& \multicolumn{5}{c}{Dice Coefficient} \\
\cmidrule(lr){2-6}          
& MED-Net & DeepLab & SegNet & FCN & UNet \\
\midrule
Background & \textbf{0.87 $\pm$ 0.03} & 0.65 $\pm$ 0.17 & 0.65 $\pm$ 0.20 & 0.86 $\pm$ 0.02 & 0.83 $\pm$ 0.06 \\
Artifact & \textbf{0.78 $\pm$ 0.04} & 0.65 $\pm$ 0.13 & 0.62 $\pm$ 0.10 &\textbf{ 0.78 $\pm$ 0.04} & 0.70 $\pm$ 0.03 \\
Meshwork & \textbf{0.58 $\pm$ 0.10} & 0.53 $\pm$ 0.13 & 0.48 $\pm$ 0.11 & 0.56 $\pm$ 0.12 & 0.30 $\pm$ 0.18 \\
Nest & 0.66 $\pm$ 0.11 & 0.55 $\pm$ 0.16 & 0.61 $\pm$ 0.13 & \textbf{0.67 $\pm$ 0.11} & 0.43 $\pm$ 0.11 \\
Ring & \textbf{0.82 $\pm$ 0.08} & 0.79 $\pm$ 0.09 & 0.78 $\pm$ 0.10 & 0.79 $\pm$ 0.11 & 0.78 $\pm$ 0.07 \\
Aspecific & \textbf{0.60 $\pm$ 0.19} & 0.25 $\pm$ 0.34 & 0.39 $\pm$ 0.23 & 0.48 $\pm$ 0.26 & 0.42 $\pm$ 0.19 \\
\bottomrule
    \end{tabular}%
    \label{tab:ResultsCV}%
\end{table*}%

\subsection{Scenario-2: Clinic-wise Cross-Validation}\label{ClinicExp}
To assess how the models generalize across clinical settings, we trained them over the data collected in Italy (86 mosaics) and tested on data collected at 4 US clinics (31 mosaics). In this case, we were not able to keep the incidences of the labels in the training and test sets at similar levels (Table~\ref{tab:DataStats}). In the training set, [18,20,21,6,23,12] percent of the labeled pixels were, [background, artifact, meshwork, nested, ring and aspecific] patterns respectively; whereas in the test set the ratios were [8,23,23,5,36,5] percent. We used the same network model architectures and training parameters that we used in Scenario~1 for both \mednet{} and the other networks.

In Table~\ref{tab:ResultsClinic}, we summarize the segmentation performance of these networks in terms of sensitivity, specificity, and Dice coefficient. In general, performances of all the networks were close to what we observed on the patient-wise stratification, with only modest decreases in the performance metrics. Overall, \mednet{} outperformed all the other networks in terms of averages across classes, particular with regards to sensitivity and Dice coefficient. Specificity values were generally very high for all networks on all classes, and for some classes, other networks had sightly higher specificity than \mednet{}.

\begin{table*}[htbp]
  \centering
  \caption{Results for Scenario 2 Clinic-Wise Cross-Validation  Experiments. The best results for each metric and label are highlighted in bold.}
  
    \begin{tabular}{rccccc}
\toprule
& \multicolumn{5}{c}{Sensitivity} \\
\cmidrule(lr){2-6} 
& MED-Net & DeepLab & SegNet & FCN & UNet \\
\midrule
    Background & 0.89  & 0.94  & \textbf{0.95}  & 0.90  & 0.77 \\
    Artifact & 0.81  & 0.69  & 0.67  & 0.78  & \textbf{0.92} \\
    Meshwork & \textbf{0.67}  & 0.57  & \textbf{0.67}  & 0.66  & 0.17 \\
    Nest & \textbf{0.50}  & 0.27  & 0.19  & 0.37  & 0.23 \\
    Ring & \textbf{0.79}  & 0.75  & 0.74  & 0.71  & 0.70 \\
    Aspecific & 0.77  & 0.72  & 0.47  & 0.65  & \textbf{0.87}\\
    \toprule
     & \multicolumn{5}{c}{Specificity} \\
\cmidrule(lr){2-6}
& MED-Net & DeepLab & SegNet & FCN & UNet \\
\midrule
    Background & \textbf{0.99}  & 0.93  & 0.90  & 0.95  & 0.91 \\
    Artifact & 0.92  & 0.93  & 0.95  & 0.91  & \textbf{0.98} \\
    Meshwork & 0.91  & 0.90  & 0.85  & 0.89  & \textbf{0.95} \\
    Nest & 0.99  & \textbf{1.00}  & \textbf{1.00}  & \textbf{1.00}  & \textbf{1.00} \\
    Ring & \textbf{0.94}  & 0.91  & 0.94  & 0.93  & \textbf{0.94} \\
    Aspecific & 0.95  & 0.95  & \textbf{0.98}  & \textbf{0.98}  & 0.79 \\
    \toprule
& \multicolumn{5}{c}{Dice Coefficient} \\

\cmidrule(lr){2-6} 
& MED-Net & DeepLab & SegNet & FCN & UNet \\
\midrule
    Background & \textbf{0.92}  & 0.85  & 0.80  & 0.86  & \textbf{0.92} \\
    Artifact &  \textbf{0.76} & 0.71  & 0.72  & 0.73  & 0.72 \\
    Meshwork & \textbf{0.67}  & 0.59  & 0.60  & 0.63  & 0.25 \\
    Nest & \textbf{0.62}  & 0.41  & 0.32  & 0.51  & 0.37 \\
    Ring &\textbf{ 0.79}  & 0.73  & 0.75  & 0.73  & 0.73 \\
    Aspecific & \textbf{0.72}  & 0.70  & 0.57  & 0.71  & 0.50 \\
    \bottomrule
    \end{tabular}
  \label{tab:ResultsClinic}%
\end{table*}%


\subsection{Ablation Studies}
We conducted 2 ablation studies to investigate how multiscale analysis and the proposed loss function each affect performance. We compared ablation results to our baseline model (the 3-level \mednet{} trained using MDSC+TV loss, see Section~\ref{ClinicExp}). We followed the same training and testing procedures in Section~\ref{ClinicExp}.

To test the effect of the multiscale approach, we trained 1-level and 2-level \mednet{} models and compared them to the 3-level \mednet{}. For a fair comparison, the number of trainable parameters for all the models is kept at 6 million. The results in Table~\ref{tab:MultiLevel} show that using the multiscale analysis improves the segmentation performance. We stopped at 3 levels because a fourth level would necessarily decrease the resolution below the size of the most of the relevant features in the images.
\begin{table*}[htbp]
  \centering
  \caption{Ablation study results for training versions of MED-Net with 3 different levels.}
    \begin{tabular}{rccccccc}
    \toprule
    & \multicolumn{7}{c}{Dice Coefficient} \\
    \cmidrule(lr){2-8}
          & Background & Artifact & Meshwork & Nest & Ring & Aspecific & \textit{Average} \\
    \midrule
    1 Level & 0.90  & 0.69  & 0.62  & 0.48  & 0.71  & 0.63  & \textit{0.67} \\
    2 Level & 0.90  & 0.70  & 0.65  & 0.52  & 0.77  & 0.63  & \textit{0.69} \\
    3 Level & \textbf{0.92} & \textbf{0.76} & \textbf{0.67} & \textbf{0.62} & \textbf{0.79} & \textbf{0.72} & \textit{\textbf{0.75}} \\
    \toprule
    & \multicolumn{7}{c}{Sensitivity} \\
    \cmidrule(lr){2-8}
    & Background & Artifact & Meshwork & Nest & Ring & Aspecific & \textit{Average} \\
    \midrule
    1 Level & 0.87  & 0.66  & \textbf{0.72}  & 0.34  & 0.67  & 0.69  & \textit{0.66} \\
    2 Level & 0.85  & 0.68  & 0.71  & 0.39  & 0.74  & 0.72  & \textit{0.68} \\
    3 Level & \textbf{0.89} & \textbf{0.81} & 0.67 & \textbf{0.50} & \textbf{0.79} & \textbf{0.77} & \textit{\textbf{0.74}} \\
    \toprule
    & \multicolumn{7}{c}{Specificity} \\
    \cmidrule(lr){2-8}
    & Background & Artifact & Meshwork & Nest & Ring & Aspecific & \textit{Average} \\
    \midrule
    1 Level & \textbf{0.99} & \textbf{0.94}  & 0.84  & \textbf{1.00} & \textbf{0.94} & 0.93  & \textit{0.94} \\
    2 Level & \textbf{0.99} & 0.93  & 0.87  & 0.99  & \textbf{0.94} & 0.92  & \textit{0.94} \\
    3 Level & \textbf{0.99} & 0.92 & \textbf{0.91} & 0.99  & \textbf{0.94} & \textbf{0.95} & \textit{\textbf{0.95}} \\
    \bottomrule
    \end{tabular}%
  \label{tab:MultiLevel}%
\end{table*}%

To test the effect of the loss function, we trained the same baseline \mednet{} model using cross-entropy, Dice loss functions, and compare the results against our MDSC+TV loss defined in Section~\ref{LossFunc}. The results in Table~\ref{tab:DiffLoss} show that using MSDC+TV as the loss function results in the best segmentation performance in terms of average Dice coefficient over all classes. 

\begin{table*}[htbp]
  \centering
  \caption{Ablation study results for training MED-Net with different loss functions.}
    \resizebox{\linewidth}{!}{
    \begin{tabular}{rccccccccc}
\toprule
& \multicolumn{3}{c}{Dice Coefficient} & \multicolumn{3}{c}{Sensitivity} & \multicolumn{3}{c}{Specificity}\\
\cmidrule(lr){2-4} \cmidrule(lr){5-7} \cmidrule(lr){8-10}
 & Cross Entropy & Dice Loss & MSDC+TV & Cross Entropy & Dice Loss & MSDC+TV & Cross Entropy & Dice Loss & MSDC+TV \\
 \midrule
Background  & 0.91  & \textbf{0.92} & \textbf{0.92} & 0.89  & \textbf{0.93} & 0.89 & 0.99  & 0.98  & \textbf{0.99}\\
Artifact & 0.78  & \textbf{0.78} & 0.76 & 0.78  & 0.74  & \textbf{0.81} & 0.95  & \textbf{0.96} & 0.92\\
Meshwork & 0.57  & 0.64  & \textbf{0.67} & 0.50  & \textbf{0.71} & 0.67 & \textbf{0.93} & 0.86  & 0.91\\
Nest  & 0.56  & 0.57  & \textbf{0.62} & \textbf{0.51}  & 0.44  & 0.50 & 0.98  & 0.99  & \textbf{0.99}\\
Ring  & 0.76  & 0.75  & \textbf{0.79} & 0.76  & 0.71  & \textbf{0.79} & 0.93  & \textbf{0.95} & 0.94 \\
Aspecific & 0.68  & 0.71  & \textbf{0.72} & \textbf{0.88}  & 0.76  & 0.77 & 0.90 & \textbf{0.95} & \textbf{0.95}\\
\textit{Average} & \textit{0.71} & \textit{0.73} & \textit{\textbf{0.75}} & \textit{0.72} & \textit{0.72} & \textit{\textbf{0.74}} &\textit{\textbf{0.95}} & \textit{\textbf{0.95}} & \textit{\textbf{0.95}}\\
    \bottomrule
    \end{tabular}
    }%
  \label{tab:DiffLoss}%
\end{table*}%

\section{Discussion and Conclusions}\label{sec:Discussion}
In this article, we present a deep-learning based semantic segmentation algorithm developed specifically for \textit{in vivo} microscopy applications other than retinal imaging. Machine-learning based analysis of in-vivo optical microscopy images has unique challenges as the textural patterns of morphology in these images are different from the patterns in natural images, and they vary extensively within classes. Hence, features developed for natural images do not generally perform well on these images. This makes deep-learning based models attractive for the analysis of microscopy images as they provide the possibility of learning the best feature representation, given an objective task. Moreover, as the deep-learning-based approaches offer ways of learning both the feature representation and the classification model in an integrated fashion, they allow greater flexibility in capturing the relationships between pixels that encode complex morphological patterns like those present in RCM images. 

Semantic segmentation also addresses another need: transparent, interpretable, machine-learning-based image analysis. Unlike diagnostic decision systems that provide a ``black box'' approach to a final diagnostic score (e.g.\nobreak\ probability of being benign or malignant)~\citep{Estreva,MelaFind,Codella2017,ISIC2016}, semantic segmentation methods provide to the user the results behind the outcomes. Thus a transparent approach can facilitate acceptance and adoption of machine learning-based approaches~\citep{EU2010}. Thus spatially-resolved, multiclass semantic segmentation algorithms such as the \mednet{} architecture proposed here have this additional advantage.

We report several promising results in this study. Although average sensitivity is moderate, specificity is very high; \mednet{} performed very well at detecting the absence of a particular pattern and did not report a lot of false positives. Hence a clinician could be highly confident about the accuracy of positive results reported by the model. Moreover, Dice coefficients of 0.73-0.75 show that the model is not only good at detecting the existence of a pattern but also successfully finds the location and the extent of the pattern. On the other hand, due to its modest sensitivity, clinicians should be aware that the model may miss patterns that are present in the data. 

Compared to the other network models that we tested, \mednet{} achieved consistently higher quantitative metrics. Among other approaches, FCN performed best and had average sensitivity, specificity, and Dice coefficient similar to \mednet{}. The qualitative results provided in Fig.~\ref{fig:exampleSeg} suggest that \mednet{} avoided inaccurately fragmented annotations. Note that both networks used the same loss function, which included an over-fragmentation penalty. Thus we conclude that this result was achieved via the multiresolution feedback mechanism introduced in the network, which provides the output of the coarser network as a prior estimate to the finer level (Fig.~\ref{fig:arch}). In this way, the model was observed to provide more coherent segmentations compared to FCN.

In the field of screening of pigmented skin lesions, \mednet{} can act as a catalyst to enable faster training of novice readers and enable the adoption of RCM screening by the wider clinical community. Initially, semantic segmentation could serve as a quality assurance layer for experts, by providing them a quantitative measure of artifacts in the collected images~\citep{KoseJID19}. Assuring to acquire images where diagnostic content is not obscured by artifacts, the expert reader can first review the images blinded to \mednet{} output, and then re-review their readings compared to an automated semantic segmentation. In this way, the semantic segmentation could offer the expert a chance to identify areas of importance that may have been missed in their initial review, and then accept or reject the \mednet{} output. Previous works have suggested that a double review of cases is preferable for remote interpretation~\citep{Witkowski17}, but this can be logistically infeasible due to the limited availability of experts. Having an integrated segmentation analysis serve as a second review may be a reasonable alternative to ensure the quality of care. In addition, \mednet{}, used together with other quantitative imaging techniques such as DEJ delineation~\citep{Kurugol2015,AlicanCVPR17,Kaur16,RobicDEJ17,AlicanML4H,HamesPlos} and diagnostic classification~\citep{Koller2011,TourneretDEJ17}, offers the potential to automate the entire image-acquisition process and pave the way for clinical imaging-based diagnostic guidance.

Although the \mednet{} was designed to work generally on microscopy images of complex tissue, we would argue for the need to be cautious when applying it directly to other domain-specific clinical microscopy applications. We needed to make domain-specific design choices in order to utilize the model and the available data to their full extent. In our case, these algorithmic choices were the minimum size of the processing area ($0.5\;\mm\times 0.5\;\mm$), the resolution of the images (2 $\mu$m/px), and the use of a multiscale CNN to increase robustness to scale changes in the morphological structures. Even if deep learning methods provide powerful solutions to represent the data of interest and carry out classification tasks, without the proper domain-specific choices, one may not achieve good results. In addition, we caution that the speckle noise inherent in optical imaging of scattering tissue poses a challenge as it changes the texture of morphological patterns and increases the variability in their appearance. In our case, we observed that designing augmentation techniques to simulate the variation in the data greatly helped in ameliorating this problem and increased both sensitivity and specificity. 

Another way to potentially increase the performance would be to increase the amount of available training data. For example, as mentioned in Section~\ref{sec:Results}, ``deformed" variants of the meshwork pattern were misclassified by \mednet{}, decreasing the segmentation performance. We believe that it is possible to overcome this problem by using more meshwork pattern that includes such deformations for training. Similar strategies could be followed to cover variations of all the patterns and increase the segmentation performance.

However, preparing data to train semantic segmentation models is logistically challenging. Unlike widely used classification models, where collecting image-wise labels are sufficient for training, data labeling for semantic segmentation is laborious and time-consuming, as it requires identifying precise and exhaustive boundaries in each test image. Additionally, unlike labeling natural scenes where the object borders are well defined, subjectivity is a common issue in labeling microscopic images. For example, even if meshwork and ring patterns are considered two different morphological patterns in their canonical form, it was not at all uncommon in our data for one of the patterns to slowly morph into the other, leading to a region with a blend of both patterns. One way to ease the experts' labeling workload, which we adopted here, was to ask experts to label only relatively clear and distinct single-pattern regions, rather than exhaustively labeling all pixels. Specifically, we asked the experts to label only the areas that they thought represented clear examples of the six given patterns. The result was that they labeled  $57\%$ of the training data pixels across the 117 mosaics. Once trained, \mednet{} was able to predict labels for the entire mosaic, although we were not able to calculate quantitative metrics on the unlabeled regions due to lack of ground truth. To allow this level of flexibility for the labelers, we designed our training procedure to be capable of handling partially labeled data by calculating and backpropagating the error over only the labeled pixels. 

However, based on our experience, we believe that even this "partial labeling"  scheme will not be sustainable in the long run if we want to significantly increase the size and variety of data available for further training and development. We are currently investigating ways of utilizing "weakly-labeled" data for semantic segmentation purposes. In such a scheme, the expert would provide only mosaic-wise labels (or maybe quadrant-wise, or for other, fixed, smaller portions of the mosaics), similar to what is done for classification problems. These labels would then be extended by the network to full semantic segmentation maps. These regions could be singly or multiply labeled according to both the ML scheme and the nature of the data. \citet{GabrieleNatMed} investigate a multiple instance learning based approach for the segmentation of histopathology slides. In histopathology, large amounts of weakly-labeled data are available through pathology slides and the respective pathology reports (e.g.\nobreak\ \citeauthor{GabrieleNatMed} used 12 thousand pathology slides). RCM imaging, on the other hand, is likely to remain in the realm of small data. We hope that this work, and specifically the availability of \mednet{}, will help to accelerate the adoption of RCM imaging, in turn leading to larger data availability in the coming years to enable the application of weakly-supervised methods. 

Finally, we wish to return to the topic of wider applicability. \mednet{} was explicitly designed as a segmentation tool that can be used for other imaging modalities and other non-generic "edge case" applications. The multiscale cellular and morphological textural patterns seen in RCM images of melanocytic skin lesions have underlying similarities to patterns seen in other tissues and conditions (e.g.\nobreak\ non-melanocytic skin lesions, skin pre-cancers, oral pre-cancers and cancers, benign and inflammatory conditions in skin~\citep{Flores19,Gary19,Longo12}) and with other emerging optical microscopic imaging approaches (optical coherence tomography (OCT), multimodal OCT-and-RCM, multiphoton microscopy (MPM), optical coherence microscopy (OCM))~\citep{Rossi19,PellacaniOCT}. Thus we also hope that the utilization of \mednet{} for both clinical training and clinical practice will eventually help to drive wider acceptance and adoption of in vivo optical microscopy in clinical practice.


\section*{Acknowledgment} 
This project was supported by NIH grant R01CA199673 from NCI and in part by MSKCC’s Cancer Center core support NIH grant P30CA008748 from NCI. The authors would like to thank NVIDIA Corporation for the Titan V GPU donation through their GPU Grant Program.

Christi Alessi-Fox is a current employee and shareholder at CaliberID. Milind Rajadhyaksha is a former employee of and holds equity in CaliberID, manufacturer of a confocal microscope. Prof. Giovanni Pellacani received honoraria for courses on confocal microscopy from Mavig GmbH, and served as advisory board member for CaliberID.

\bibliographystyle{abbrvnat}
\bibliography{RCM}

\begin{thebibliography}{59}
\providecommand{\natexlab}[1]{#1}
\providecommand{\url}[1]{\texttt{#1}}
\expandafter\ifx\csname urlstyle\endcsname\relax
  \providecommand{\doi}[1]{doi: #1}\else
  \providecommand{\doi}{doi: \begingroup \urlstyle{rm}\Url}\fi

\bibitem[Alarcon et~al.(2014)Alarcon, Carrera, Palou, Alos, Malvehy, and
  Puig]{Alarcon14}
I.~Alarcon, C.~Carrera, J.~Palou, L.~Alos, J.~Malvehy, and S.~Puig.
\newblock Impact of in vivo reflectance confocal microscopy on the number
  needed to treat melanoma in doubtful lesions.
\newblock \emph{British journal of Dermatology}, 170\penalty0 (4):\penalty0
  802--808, 2014.

\bibitem[Amirul~Islam et~al.(2017)Amirul~Islam, Rochan, Bruce, and
  Wang]{Islam17}
M.~Amirul~Islam, M.~Rochan, N.~D. Bruce, and Y.~Wang.
\newblock Gated feedback refinement network for dense image labeling.
\newblock In \emph{Proceedings of the IEEE Conference on Computer Vision and
  Pattern Recognition}, pages 3751--3759, 2017.

\bibitem[Badrinarayanan et~al.(2017)Badrinarayanan, Kendall, and
  Cipolla]{segnet}
V.~Badrinarayanan, A.~Kendall, and R.~Cipolla.
\newblock Segnet: A deep convolutional encoder-decoder architecture for image
  segmentation.
\newblock \emph{IEEE transactions on pattern analysis and machine
  intelligence}, 39\penalty0 (12):\penalty0 2481--2495, 2017.

\bibitem[Boone et~al.(2015)Boone, Marneffe, Suppa, Miyamoto, Alarcon,
  Hofmann-Wellenhof, Malvehy, Pellacani, and Del~Marmol]{PellacaniOCT}
M.~Boone, A.~Marneffe, M.~Suppa, M.~Miyamoto, I.~Alarcon, R.~Hofmann-Wellenhof,
  J.~Malvehy, G.~Pellacani, and V.~Del~Marmol.
\newblock High-definition optical coherence tomography algorithm for the
  discrimination of actinic keratosis from normal skin and from squamous cell
  carcinoma.
\newblock \emph{Journal of the European Academy of Dermatology and
  Venereology}, 29\penalty0 (8):\penalty0 1606--1615, 2015.

\bibitem[Borsari et~al.(2016)Borsari, Pampena, Lallas, Kyrgidis, Moscarella,
  Benati, Raucci, Pellacani, Zalaudek, Argenziano, et~al.]{Borsari16}
S.~Borsari, R.~Pampena, A.~Lallas, A.~Kyrgidis, E.~Moscarella, E.~Benati,
  M.~Raucci, G.~Pellacani, I.~Zalaudek, G.~Argenziano, et~al.
\newblock Clinical indications for use of reflectance confocal microscopy for
  skin cancer diagnosis.
\newblock \emph{JAMA dermatology}, 152\penalty0 (10):\penalty0 1093--1098,
  2016.

\bibitem[Bozkurt et~al.(2017{\natexlab{a}})Bozkurt, Gale, Kose, Alessi-Fox,
  Brooks, Rajadhyaksha, and Dy]{AlicanCVPR17}
A.~Bozkurt, T.~Gale, K.~Kose, C.~Alessi-Fox, D.~H. Brooks, M.~Rajadhyaksha, and
  J.~Dy.
\newblock Delineation of skin strata in reflectance confocal microscopy images
  with recurrent convolutional networks.
\newblock In \emph{Proceedings of the IEEE Conference on Computer Vision and
  Pattern Recognition Workshops}, pages 25--33, 2017{\natexlab{a}}.

\bibitem[Bozkurt et~al.(2017{\natexlab{b}})Bozkurt, Kose, Coll-Font,
  Alessi-Fox, Brooks, Dy, and Rajadhyaksha]{AlicanML4H}
A.~Bozkurt, K.~Kose, J.~Coll-Font, C.~Alessi-Fox, D.~H. Brooks, J.~G. Dy, and
  M.~Rajadhyaksha.
\newblock Delineation of skin strata in reflectance confocal microscopy images
  using recurrent convolutional networks with toeplitz attention.
\newblock \emph{arXiv preprint arXiv:1712.00192}, 2017{\natexlab{b}}.

\bibitem[Bozkurt et~al.(2018)Bozkurt, Kose, Alessi-Fox, Gill, Dy, Brooks, and
  Rajadhyaksha]{KoseM18}
A.~Bozkurt, K.~Kose, C.~Alessi-Fox, M.~Gill, J.~Dy, D.~Brooks, and
  M.~Rajadhyaksha.
\newblock A multiresolution convolutional neural network with partial label
  training for annotating reflectance confocal microscopy images of skin.
\newblock In \emph{International Conference on Medical Image Computing and
  Computer-Assisted Intervention}, pages 292--299. Springer, 2018.

\bibitem[Campanella et~al.(2019)Campanella, Hanna, Geneslaw, Miraflor, Silva,
  Busam, Brogi, Reuter, Klimstra, and Fuchs]{GabrieleNatMed}
G.~Campanella, M.~G. Hanna, L.~Geneslaw, A.~Miraflor, V.~W.~K. Silva, K.~J.
  Busam, E.~Brogi, V.~E. Reuter, D.~S. Klimstra, and T.~J. Fuchs.
\newblock Clinical-grade computational pathology using weakly supervised deep
  learning on whole slide images.
\newblock \emph{Nature medicine}, 25\penalty0 (8):\penalty0 1301--1309, 2019.

\bibitem[Chen et~al.(2016)Chen, Papandreou, Kokkinos, Murphy, and
  Yuille]{deeplab}
L.-C. Chen, G.~Papandreou, I.~Kokkinos, K.~Murphy, and A.~L. Yuille.
\newblock Deeplab: Semantic image segmentation with deep convolutional nets,
  atrous convolution, and fully connected crfs.
\newblock \emph{arXiv preprint arXiv:1606.00915}, 2016.

\bibitem[Chen et~al.(2018)Chen, Gadepalli, MacDonald, Liu, Nagpal, Kohlberger,
  Dean, Corrado, Hipp, and Stumpe]{GooglePath}
P.-H.~C. Chen, K.~Gadepalli, R.~MacDonald, Y.~Liu, K.~Nagpal, T.~Kohlberger,
  J.~Dean, G.~S. Corrado, J.~D. Hipp, and M.~C. Stumpe.
\newblock Microscope 2.0: An augmented reality microscope with real-time
  artificial intelligence integration.
\newblock \emph{arXiv preprint arXiv:1812.00825}, 2018.

\bibitem[CIBC(2016)]{Seg3D}
CIBC, 2016.
\newblock Seg3D: Volumetric Image Segmentation and Visualization. Scientific
  Computing and Imaging Institute (SCI), Download from: http://www.seg3d.org.

\bibitem[Codella et~al.(2017)Codella, Nguyen, Pankanti, Gutman, Helba, Halpern,
  and Smith]{Codella2017}
N.~C. Codella, Q.-B. Nguyen, S.~Pankanti, D.~A. Gutman, B.~Helba, A.~C.
  Halpern, and J.~R. Smith.
\newblock Deep learning ensembles for melanoma recognition in dermoscopy
  images.
\newblock \emph{IBM Journal of Research and Development}, 61\penalty0
  (4/5):\penalty0 5--1, 2017.

\bibitem[Codella et~al.(2018)Codella, Gutman, Celebi, Helba, Marchetti, Dusza,
  Kalloo, Liopyris, Mishra, Kittler, et~al.]{ISIC2017}
N.~C. Codella, D.~Gutman, M.~E. Celebi, B.~Helba, M.~A. Marchetti, S.~W. Dusza,
  A.~Kalloo, K.~Liopyris, N.~Mishra, H.~Kittler, et~al.
\newblock Skin lesion analysis toward melanoma detection: A challenge at the
  2017 international symposium on biomedical imaging ({ISBI}), hosted by the
  international skin imaging collaboration {(ISIC)}.
\newblock In \emph{2018 IEEE 15th International Symposium on Biomedical Imaging
  (ISBI 2018)}, pages 168--172. IEEE, 2018.

\bibitem[Dice(1945)]{dice1945measures}
L.~R. Dice.
\newblock Measures of the amount of ecologic association between species.
\newblock \emph{Ecology}, 26\penalty0 (3):\penalty0 297--302, 1945.

\bibitem[Eigen and Fergus(2015)]{Fergus15}
D.~Eigen and R.~Fergus.
\newblock Predicting depth, surface normals and semantic labels with a common
  multi-scale convolutional architecture.
\newblock In \emph{Proceedings of the IEEE international conference on computer
  vision}, pages 2650--2658, 2015.

\bibitem[Esteva et~al.(2017)Esteva, Kuprel, Novoa, Ko, Swetter, Blau, and
  Thrun]{Estreva}
A.~Esteva, B.~Kuprel, R.~A. Novoa, J.~Ko, S.~M. Swetter, H.~M. Blau, and
  S.~Thrun.
\newblock {Dermatologist-level classification of skin cancer with deep neural
  networks}.
\newblock \emph{Nature}, 542\penalty0 (7639):\penalty0 115--118, Jan. 2017.
\newblock ISSN 0028-0836.
\newblock \doi{10.1038/nature21056}.

\bibitem[Falk et~al.(2019)Falk, Mai, Bensch, {\c{C}}i{\c{c}}ek, Abdulkadir,
  Marrakchi, B{\"o}hm, Deubner, J{\"a}ckel, Seiwald, et~al.]{unet19}
T.~Falk, D.~Mai, R.~Bensch, {\"O}.~{\c{C}}i{\c{c}}ek, A.~Abdulkadir,
  Y.~Marrakchi, A.~B{\"o}hm, J.~Deubner, Z.~J{\"a}ckel, K.~Seiwald, et~al.
\newblock U-net: deep learning for cell counting, detection, and morphometry.
\newblock \emph{Nature methods}, 16\penalty0 (1):\penalty0 67, 2019.

\bibitem[Flores et~al.(2019)Flores, Y{\'e}lamos, Cordova, Kose, Phillips, Lee,
  Rossi, Nehal, and Rajadhyaksha]{Flores19}
E.~Flores, O.~Y{\'e}lamos, M.~Cordova, K.~Kose, W.~Phillips, E.~Lee, A.~Rossi,
  K.~Nehal, and M.~Rajadhyaksha.
\newblock Peri-operative delineation of non-melanoma skin cancer margins in
  vivo with handheld reflectance confocal microscopy and video-mosaicking.
\newblock \emph{Journal of the European Academy of Dermatology and
  Venereology}, 33\penalty0 (6):\penalty0 1084--1091, 2019.

\bibitem[Fu et~al.(2016)Fu, Xu, Wong, and Liu]{FuRetina16}
H.~Fu, Y.~Xu, D.~W.~K. Wong, and J.~Liu.
\newblock Retinal vessel segmentation via deep learning network and
  fully-connected conditional random fields.
\newblock In \emph{2016 IEEE 13th international symposium on biomedical imaging
  (ISBI)}, pages 698--701. IEEE, 2016.

\bibitem[Fu et~al.(2018)Fu, Cheng, Xu, Wong, Liu, and Cao]{Fu18}
H.~Fu, J.~Cheng, Y.~Xu, D.~W.~K. Wong, J.~Liu, and X.~Cao.
\newblock Joint optic disc and cup segmentation based on multi-label deep
  network and polar transformation.
\newblock \emph{IEEE transactions on medical imaging}, 37\penalty0
  (7):\penalty0 1597--1605, 2018.

\bibitem[Gill et~al.(2019)Gill, Alessi-Fox, and Kose]{Artifacts}
M.~Gill, C.~Alessi-Fox, and K.~Kose.
\newblock Artifacts and landmarks: pearls and pitfalls for in vivo reflectance
  confocal microscopy of the skin using the tissue-coupled device.
\newblock \emph{Dermatology online journal}, 25\penalty0 (8), 2019.

\bibitem[Goodman and Flaxman(2017)]{EU2010}
B.~Goodman and S.~Flaxman.
\newblock European union regulations on algorithmic decision-making and a
  ``right to explanation''.
\newblock \emph{AI Magazine}, 38\penalty0 (3):\penalty0 50--57, 2017.

\bibitem[Gu et~al.(2018)Gu, Burlutskiy, Andersson, and Wil{\'e}n]{Feng18}
F.~Gu, N.~Burlutskiy, M.~Andersson, and L.~K. Wil{\'e}n.
\newblock Multi-resolution networks for semantic segmentation in whole slide
  images.
\newblock In \emph{Computational Pathology and Ophthalmic Medical Image
  Analysis}, pages 11--18. Springer, 2018.

\bibitem[Gu et~al.(2019)Gu, Cheng, Fu, Zhou, Hao, Zhao, Zhang, Gao, and
  Liu]{Gu19}
Z.~Gu, J.~Cheng, H.~Fu, K.~Zhou, H.~Hao, Y.~Zhao, T.~Zhang, S.~Gao, and J.~Liu.
\newblock Ce-net: Context encoder network for 2d medical image segmentation.
\newblock \emph{IEEE transactions on medical imaging}, 2019.

\bibitem[Guy~Jr et~al.(2015)Guy~Jr, Machlin, Ekwueme, and
  Yabroff]{SkinCancerCost}
G.~P. Guy~Jr, S.~R. Machlin, D.~U. Ekwueme, and K.~R. Yabroff.
\newblock Prevalence and costs of skin cancer treatment in the us, 2002- 2006
  and 2007- 2011.
\newblock \emph{American journal of preventive medicine}, 48\penalty0
  (2):\penalty0 183--187, 2015.

\bibitem[Halimi et~al.(2017)Halimi, Batatia, Jimmy, Josse, and
  Tourneret]{TourneretDEJ17}
A.~Halimi, H.~Batatia, L.~D. Jimmy, G.~Josse, and J.~Y. Tourneret.
\newblock Wavelet-based statistical classification of skin images acquired with
  reflectance confocal microscopy.
\newblock \emph{Biomedical Optics Express}, 8\penalty0 (12):\penalty0
  5450--5467, 2017.

\bibitem[Hames et~al.(2016)Hames, Ardig{\`o}, Soyer, Bradley, and
  Prow]{HamesPlos}
S.~C. Hames, M.~Ardig{\`o}, H.~P. Soyer, A.~P. Bradley, and T.~W. Prow.
\newblock Automated segmentation of skin strata in reflectance confocal
  microscopy depth stacks.
\newblock \emph{PloS one}, 11\penalty0 (4):\penalty0 e0153208, 2016.

\bibitem[He et~al.(2016)He, Zhang, Ren, and Sun]{HeNormal}
K.~He, X.~Zhang, S.~Ren, and J.~Sun.
\newblock Deep residual learning for image recognition.
\newblock In \emph{Proceedings of the IEEE conference on computer vision and
  pattern recognition}, pages 770--778, 2016.

\bibitem[Jiang et~al.(2018)Jiang, Hu, Liu, Halpenny, Hellmann, Deasy, Mageras,
  and Veeraraghavan]{Jiang18}
J.~Jiang, Y.-C. Hu, C.-J. Liu, D.~Halpenny, M.~D. Hellmann, J.~O. Deasy,
  G.~Mageras, and H.~Veeraraghavan.
\newblock Multiple resolution residually connected feature streams for
  automatic lung tumor segmentation from ct images.
\newblock \emph{IEEE transactions on medical imaging}, 38\penalty0
  (1):\penalty0 134--144, 2018.

\bibitem[Kaur et~al.(2016)Kaur, Dana, Cula, and Mack]{Kaur16}
P.~Kaur, K.~J. Dana, G.~O. Cula, and M.~C. Mack.
\newblock Hybrid deep learning for reflectance confocal microscopy skin images.
\newblock In \emph{2016 23rd International Conference on Pattern Recognition
  (ICPR)}, pages 1466--1471. IEEE, 2016.

\bibitem[Koller et~al.(2011)Koller, Wiltgen, Ahlgrimm~Siess, Weger,
  Hofmann~Wellenhof, Richtig, Smolle, and Gerger]{Koller2011}
S.~Koller, M.~Wiltgen, V.~Ahlgrimm~Siess, W.~Weger, R.~Hofmann~Wellenhof,
  E.~Richtig, J.~Smolle, and A.~Gerger.
\newblock In vivo reflectance confocal microscopy: automated diagnostic image
  analysis of melanocytic skin tumours.
\newblock \emph{Journal of European Academy of Dermatology and Venerology},
  25\penalty0 (5):\penalty0 5, 2011.

\bibitem[Kose et~al.(2019)Kose, Bozkurt, Alessi-Fox, Brooks, Dy, Rajadhyaksha,
  and Gill]{KoseJID19}
K.~Kose, A.~Bozkurt, C.~Alessi-Fox, D.~H. Brooks, J.~G. Dy, M.~Rajadhyaksha,
  and M.~Gill.
\newblock Utilizing machine learning for image quality assessment for
  reflectance confocal microscopy.
\newblock \emph{Journal of Investigative Dermatology}, 2019.
\newblock ISSN 0022-202X.
\newblock \doi{https://doi.org/10.1016/j.jid.2019.10.018}.

\bibitem[Kurugol et~al.(2015)Kurugol, Kose, Park, Dy, Brooks, and
  Rajadhyaksha]{Kurugol2015}
S.~Kurugol, K.~Kose, B.~Park, J.~G. Dy, D.~H. Brooks, and M.~Rajadhyaksha.
\newblock Automated delineation of dermal--epidermal junction in reflectance
  confocal microscopy image stacks of human skin.
\newblock \emph{Journal of Investigative Dermatology}, 135\penalty0
  (3):\penalty0 710--717, 2015.

\bibitem[Li et~al.(2017)Li, Sarma, Ho, Gertych, Knudsen, and Arnold]{Sarma18}
J.~Li, K.~V. Sarma, K.~C. Ho, A.~Gertych, B.~S. Knudsen, and C.~W. Arnold.
\newblock A multi-scale u-net for semantic segmentation of histological images
  from radical prostatectomies.
\newblock In \emph{AMIA Annual Symposium Proceedings}, volume 2017, page 1140.
  American Medical Informatics Association, 2017.

\bibitem[Lin et~al.(2017)Lin, Milan, Shen, and Reid]{refinenet}
G.~Lin, A.~Milan, C.~Shen, and I.~Reid.
\newblock Refinenet: Multi-path refinement networks for high-resolution
  semantic segmentation.
\newblock In \emph{Proceedings of the IEEE conference on computer vision and
  pattern recognition}, pages 1925--1934, 2017.

\bibitem[Litjens et~al.(2017)Litjens, Kooi, Bejnordi, Setio, Ciompi,
  Ghafoorian, Van Der~Laak, Van~Ginneken, and S{\'a}nchez]{Litjens17}
G.~Litjens, T.~Kooi, B.~E. Bejnordi, A.~A.~A. Setio, F.~Ciompi, M.~Ghafoorian,
  J.~A. Van Der~Laak, B.~Van~Ginneken, and C.~I. S{\'a}nchez.
\newblock A survey on deep learning in medical image analysis.
\newblock \emph{Medical image analysis}, 42:\penalty0 60--88, 2017.

\bibitem[Long et~al.(2015)Long, Shelhamer, and Darrell]{fcn}
J.~Long, E.~Shelhamer, and T.~Darrell.
\newblock Fully convolutional networks for semantic segmentation.
\newblock In \emph{Proceedings of the IEEE conference on computer vision and
  pattern recognition}, pages 3431--3440, 2015.

\bibitem[Longo et~al.(2012)Longo, Zalaudek, Argenziano, and Pellacani]{Longo12}
C.~Longo, I.~Zalaudek, G.~Argenziano, and G.~Pellacani.
\newblock New directions in dermatopathology: in vivo confocal microscopy in
  clinical practice.
\newblock \emph{Dermatologic clinics}, 30\penalty0 (4):\penalty0 799--814,
  2012.

\bibitem[Marchetti et~al.(2018)Marchetti, Codella, Dusza, Gutman, Helba,
  Kalloo, Mishra, Carrera, Celebi, DeFazio, et~al.]{ISIC2016}
M.~A. Marchetti, N.~C. Codella, S.~W. Dusza, D.~A. Gutman, B.~Helba, A.~Kalloo,
  N.~Mishra, C.~Carrera, M.~E. Celebi, J.~L. DeFazio, et~al.
\newblock Results of the 2016 international skin imaging collaboration
  international symposium on biomedical imaging challenge: Comparison of the
  accuracy of computer algorithms to dermatologists for the diagnosis of
  melanoma from dermoscopic images.
\newblock \emph{Journal of the American Academy of Dermatology}, 78\penalty0
  (2):\penalty0 270--277, 2018.

\bibitem[Monheit et~al.(2011)Monheit, Cognetta, Ferris, Rabinovitz, Gross,
  Martini, Grichnik, Mihm, Prieto, Googe, et~al.]{MelaFind}
G.~Monheit, A.~B. Cognetta, L.~Ferris, H.~Rabinovitz, K.~Gross, M.~Martini,
  J.~M. Grichnik, M.~Mihm, V.~G. Prieto, P.~Googe, et~al.
\newblock The performance of melafind: a prospective multicenter study.
\newblock \emph{Archives of dermatology}, 147\penalty0 (2):\penalty0 188--194,
  2011.

\bibitem[Nie et~al.(2016)Nie, Wang, Gao, and Shen]{CT-MR-FCN}
D.~Nie, L.~Wang, Y.~Gao, and D.~Shen.
\newblock Fully convolutional networks for multi-modality isointense infant
  brain image segmentation.
\newblock In \emph{2016 IEEE 13Th international symposium on biomedical imaging
  (ISBI)}, pages 1342--1345. IEEE, 2016.

\bibitem[Nikolaou and Stratigos(2014)]{nikolaou2014emerging}
V.~Nikolaou and A.~Stratigos.
\newblock Emerging trends in the epidemiology of melanoma.
\newblock \emph{British journal of dermatology}, 170\penalty0 (1):\penalty0
  11--19, 2014.

\bibitem[Pellacani et~al.(2014)Pellacani, Pepe, Casari, and Longo]{Pellacani14}
G.~Pellacani, P.~Pepe, A.~Casari, and C.~Longo.
\newblock Reflectance confocal microscopy as a second-level examination in skin
  oncology improves diagnostic accuracy and saves unnecessary excisions: a
  longitudinal prospective study.
\newblock \emph{British Journal of Dermatology}, 171\penalty0 (5):\penalty0
  1044--1051, 2014.

\bibitem[Pellacani et~al.(2016)Pellacani, Witkowski, Cesinaro, Losi, Colombo,
  Campagna, Longo, Piana, De~Carvalho, Giusti, et~al.]{Pellacani16}
G.~Pellacani, A.~Witkowski, A.~Cesinaro, A.~Losi, G.~Colombo, A.~Campagna,
  C.~Longo, S.~Piana, N.~De~Carvalho, F.~Giusti, et~al.
\newblock Cost--benefit of reflectance confocal microscopy in the diagnostic
  performance of melanoma.
\newblock \emph{Journal of the European Academy of Dermatology and
  Venereology}, 30\penalty0 (3):\penalty0 413--419, 2016.

\bibitem[Peterson et~al.(2019)Peterson, Zanoni, Ardigo, Migliacci, Patel, and
  Rajadhyaksha]{Gary19}
G.~Peterson, D.~K. Zanoni, M.~Ardigo, J.~C. Migliacci, S.~G. Patel, and
  M.~Rajadhyaksha.
\newblock Feasibility of a video-mosaicking approach to extend the
  field-of-view for reflectance confocal microscopy in the oral cavity in vivo.
\newblock \emph{Lasers in Surgery and Medicine}, 51\penalty0 (5):\penalty0
  439--451, 2019.

\bibitem[Rajadhyaksha et~al.(2017)Rajadhyaksha, Marghoob, Rossi, Halpern, and
  Nehal]{rajadhyaksha2017reflectance}
M.~Rajadhyaksha, A.~Marghoob, A.~Rossi, A.~C. Halpern, and K.~S. Nehal.
\newblock Reflectance confocal microscopy of skin in vivo: From bench to
  bedside.
\newblock \emph{Lasers in surgery and medicine}, 49\penalty0 (1):\penalty0
  7--19, 2017.

\bibitem[Robic et~al.(2017)Robic, Perret, Nkegne, Couprie, and
  Talbot]{RobicDEJ17}
J.~Robic, B.~Perret, A.~Nkegne, M.~Couprie, and H.~Talbot.
\newblock Classification of the dermal-epidermal junction using in-vivo
  confocal microscopy.
\newblock In \emph{2017 IEEE 14th International Symposium on Biomedical Imaging
  (ISBI 2017)}, pages 252--255, April 2017.
\newblock \doi{10.1109/ISBI.2017.7950513}.

\bibitem[Ronneberger et~al.(2015)Ronneberger, Fischer, and
  Brox]{ronneberger2015u}
O.~Ronneberger, P.~Fischer, and T.~Brox.
\newblock U-net: Convolutional networks for biomedical image segmentation.
\newblock In \emph{International Conference on Medical image computing and
  computer-assisted intervention}, pages 234--241. Springer, 2015.

\bibitem[Roy et~al.(2018)Roy, Navab, and Wachinger]{SqueezeExcite}
A.~G. Roy, N.~Navab, and C.~Wachinger.
\newblock Recalibrating fully convolutional networks with spatial and channel
  “squeeze and excitation” blocks.
\newblock \emph{IEEE transactions on medical imaging}, 38\penalty0
  (2):\penalty0 540--549, 2018.

\bibitem[Salehi et~al.(2017)Salehi, Erdogmus, and Gholipour]{Salehi17}
S.~S.~M. Salehi, D.~Erdogmus, and A.~Gholipour.
\newblock Tversky loss function for image segmentation using 3d fully
  convolutional deep networks.
\newblock In \emph{International Workshop on Machine Learning in Medical
  Imaging}, pages 379--387. Springer, 2017.

\bibitem[Schneider et~al.(2019)Schneider, Kohli, Hamzavi, Council, Rossi, and
  Ozog]{Rossi19}
S.~L. Schneider, I.~Kohli, I.~H. Hamzavi, M.~L. Council, A.~M. Rossi, and D.~M.
  Ozog.
\newblock Emerging imaging technologies in dermatology: Part ii: Applications
  and limitations.
\newblock \emph{Journal of the American Academy of Dermatology}, 80\penalty0
  (4):\penalty0 1121--1131, 2019.

\bibitem[Scope et~al.(2017)Scope, Guitera, and Pellacani]{ScopeTermChap17}
A.~Scope, P.~Guitera, and G.~Pellacani.
\newblock Rcm diagnosis of melanocytic neoplasms: Terminology, algorithms and
  their accuracy and clinical integration.
\newblock In S.~Gonz{\'a}lez, M.~Rajadhyaksha, M.~Ardigo, C.~Longo, C.~Carrera,
  M.~Ulrich, and E.~Moscarella, editors, \emph{Reflectance Confocal Microscopy
  of Cutaneous Tumors, 2nd Ed}, pages 168--186. Boca Raton, CRC Press, 2017.

\bibitem[Witkowski et~al.(2017)Witkowski, {\L}udzik, Arginelli, Bassoli,
  Benati, Casari, De~Carvalho, De~Pace, Farnetani, Losi, et~al.]{Witkowski17}
A.~Witkowski, J.~{\L}udzik, F.~Arginelli, S.~Bassoli, E.~Benati, A.~Casari,
  N.~De~Carvalho, B.~De~Pace, F.~Farnetani, A.~Losi, et~al.
\newblock Improving diagnostic sensitivity of combined dermoscopy and
  reflectance confocal microscopy imaging through double reader concordance
  evaluation in telemedicine settings: A retrospective study of 1000 equivocal
  cases.
\newblock \emph{PloS one}, 12\penalty0 (11):\penalty0 e0187748, 2017.

\bibitem[Yu et~al.(2017)Yu, Yang, Chen, Qin, and Heng]{Prostate2017}
L.~Yu, X.~Yang, H.~Chen, J.~Qin, and P.-A. Heng.
\newblock Volumetric convnets with mixed residual connections for automated
  prostate segmentation from 3d mr images.
\newblock In \emph{AAAI}, pages 66--72, 2017.

\bibitem[Zhang et~al.(2019)Zhang, Fu, Yan, Zhang, Wu, Yang, Tan, and Xu]{Fu19}
S.~Zhang, H.~Fu, Y.~Yan, Y.~Zhang, Q.~Wu, M.~Yang, M.~Tan, and Y.~Xu.
\newblock Attention guided network for retinal image segmentation.
\newblock In \emph{International Conference on Medical Image Computing and
  Computer-Assisted Intervention}, pages 797--805. Springer, 2019.

\bibitem[Zhao et~al.(2017)Zhao, Shi, Qi, Wang, and Jia]{pspnet}
H.~Zhao, J.~Shi, X.~Qi, X.~Wang, and J.~Jia.
\newblock Pyramid scene parsing network.
\newblock In \emph{Proceedings of the IEEE conference on computer vision and
  pattern recognition}, pages 2881--2890, 2017.

\bibitem[Zhou et~al.(2018)Zhou, Siddiquee, Tajbakhsh, and Liang]{Zhou18}
Z.~Zhou, M.~M.~R. Siddiquee, N.~Tajbakhsh, and J.~Liang.
\newblock Unet++: A nested u-net architecture for medical image segmentation.
\newblock In \emph{Deep Learning in Medical Image Analysis and Multimodal
  Learning for Clinical Decision Support}, pages 3--11. Springer, 2018.

\bibitem[Zhu et~al.(2017)Zhu, Du, Turkbey, Choyke, and Yan]{zhu2017deeply}
Q.~Zhu, B.~Du, B.~Turkbey, P.~L. Choyke, and P.~Yan.
\newblock Deeply-supervised cnn for prostate segmentation.
\newblock In \emph{Neural Networks (IJCNN), 2017 International Joint Conference
  on}, pages 178--184. IEEE, 2017.

\end{thebibliography}


\end{document}